\documentclass[a4paper,prl,superscriptaddress,aps,notitlepage,floatfix, twocolumn]{revtex4-2}
\usepackage[utf8]{inputenc}
\usepackage[T1]{fontenc}
\usepackage[normalem]{ulem}
\usepackage{amsmath, amssymb, ulem, array}
\usepackage[usenames]{color}
\usepackage{graphicx}
\usepackage{hhline,layout}
\usepackage{siunitx}

\usepackage[pdftex,unicode,colorlinks,citecolor=blue,linkcolor=blue,
bookmarks=true,bookmarksopen=true,bookmarksopenlevel=3,bookmarksnumbered=true]{hyperref}
\usepackage[capitalize]{cleveref}

\newcommand{\figref}[1]{Fig.~\ref{fig:#1}}

\begin{document}

\title{Fundamental sensitivity limit of lossy cavity-enhanced interferometers with external and internal squeezing}
\author{M. Korobko}
\affiliation{Institut f\"ur Laserphysik und Zentrum f\"ur Optische Quantentechnologien der Universit\"at Hamburg,\\%
Luruper Chaussee 149, 22761 Hamburg, Germany}
\author{J. S\"udbeck}
\affiliation{Institut f\"ur Laserphysik und Zentrum f\"ur Optische Quantentechnologien der Universit\"at Hamburg,\\%
Luruper Chaussee 149, 22761 Hamburg, Germany}
\author{S. Steinlechner}
\affiliation{ Faculty of Science and Engineering, Maastricht University, Duboisdomein 30, 6229 GT Maastricht, The Netherlands}
\affiliation{ Nikhef, Science Park 105, 1098 XG Amsterdam, The Netherlands }
\author{R. Schnabel}
\affiliation{Institut f\"ur Laserphysik und Zentrum f\"ur Optische Quantentechnologien der Universit\"at Hamburg,\\%
Luruper Chaussee 149, 22761 Hamburg, Germany}
\date{\today}
\begin{abstract}
Quantum optical sensors are ubiquitous in various fields of research, from biological or medical sensors to large-scale experiments searching for dark matter or gravitational waves. 
Gravitational-wave detectors have been very successful in implementing cavities and quantum squeezed light for enhancing sensitivity to signals from black hole or neutron star mergers.
However, the sensitivity to weak forces is limited by available energy and optical decoherence in the system. 
Here, we derive the fundamental sensitivity limit of cavity and squeezed-light enhanced interferometers with optical loss.
This limit is attained by the optimal use of an additional internal squeeze operation, which allows to mitigate readout loss.
We demonstrate the application of internal squeezing to various scenarios and confirm that it indeed allows to reach the best sensitivity in cavity and squeezed-light enhanced linear force sensors.
Our work establishes the groundwork for the future development of optimal sensors in real-world scenarios where, up until now, the application of squeezed light was curtailed by various sources of decoherence.
\end{abstract}

\maketitle
\section{Introduction}
Laser interferometers serve a multiple of applications, including biological macromolecule sensors \cite{Young2018}, medical sensors \cite{Nolte2012}, accelerometers \cite{Krause2012}, and extraordinary sensitive gravitational-wave detection \cite{GW150914}.
Highest sensitivities in laser interferometry are accomplished by injecting a quasi-monochromatic carrier light in a significantly displaced coherent state alongside strongly squeezed vacuum states that cover the signal band's sideband frequencies \cite{Caves1981, Schnabel2017}. 
This process is complemented by utilizing optical cavities along the interferometer arms \cite{Meers1988,Mizuno1993}, and reducing light decoherence\cite{Dorner2009,Demkowicz-Dobrzanski2013}. 
Ideally, the power of the carrier light is enhanced up to a threshold at which either the measured sample gets disturbed or the  interferometer's optical hardware compromises its quality. 
The cavities resonantly enhance both the power of the carrier light as well as the signal due to constructive interference over the many cavity round trips. 
Preceding and current gravitational-wave observatories employ the highest power that the interferometer can sustain, enhancement cavities, and externally produced squeezed vacuum states \cite{LSC2011,Grote2013,Tse2019,Acernese2019,Yu2020,Lough2021}. 
Laser interferometers have been enhanced in proof-of-principle experiments for biological sensors \cite{Taylor2013, Taylor2016}, magnetic sensors~\cite{Li2018}, optomechanical sensors \cite{Yap2020,Pooser2020}, and dark matter sensors~\cite{backesQuantumEnhancedSearch2021,Carney2021}.
However, decoherence limits the quantum enhancement provided by squeezed light, stemming from optical loss that incorporates imperfect photo-electric detection efficiency and various propagation losses. 
Decoherence always reduces the sensitivity that can be achieved.

\begin{figure}
\includegraphics[width=0.8\columnwidth]{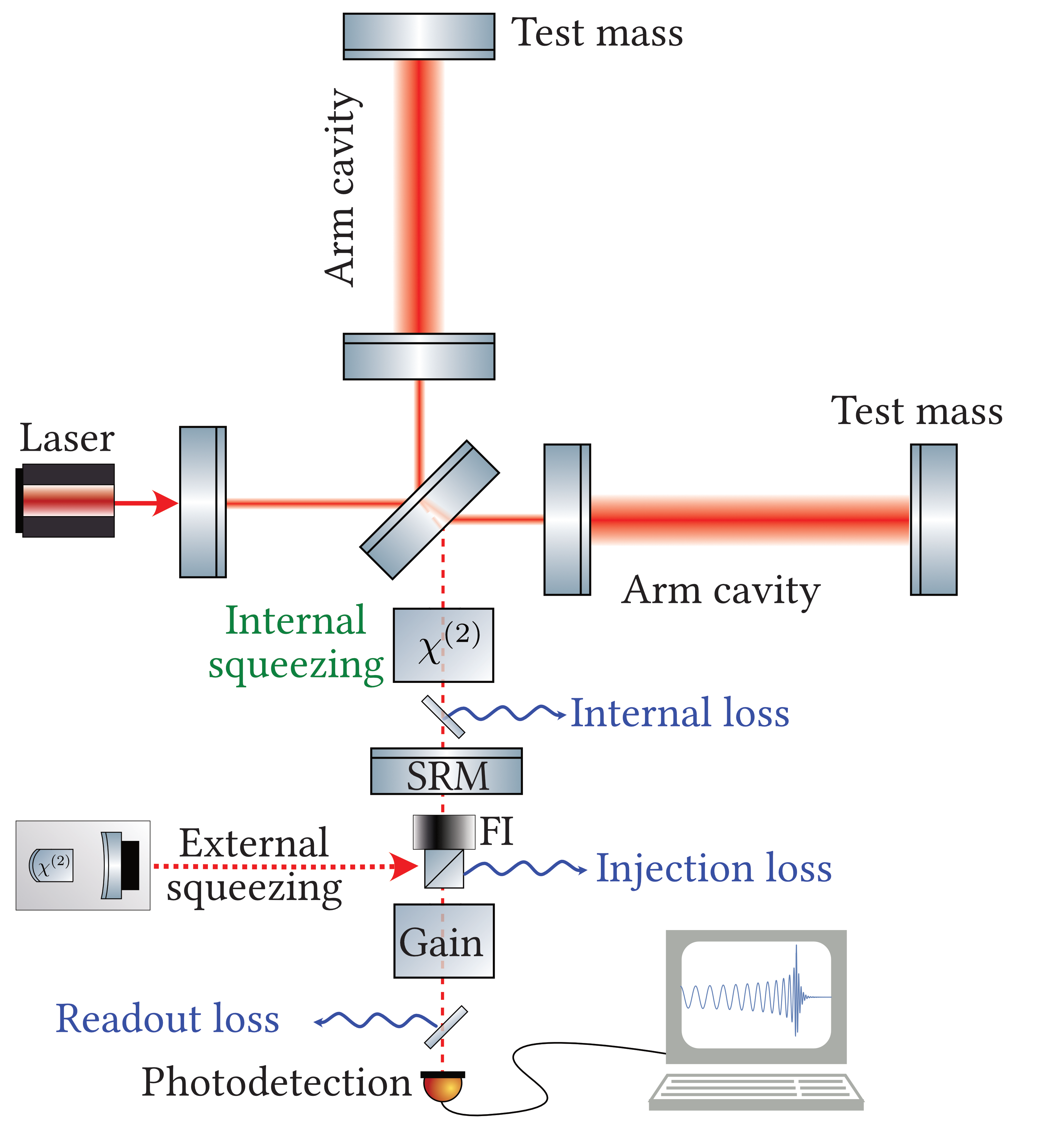}
\caption{Illustration of a future gravitational-wave detector able to operate at the fundamental sensitivity limit.
Optical losses occur inside the detectors's cavities formed by the signal-recycling mirror (SRM) and the arm cavities; upon readout, including the propagation losses, e.g. through the Faraday isolator (FI) used to inject external squeezing, as well as the loss at the photodetectors.
Injected squeezing also experiences injection loss due to various generation and propagation inefficiencies.
Internal squeezing is crucial in achieving the fundamental sensitivity limit in this detector.
All three squeeze operation: internal, external and output amplification are required in the general case.
}\label{fig_gwsetup}
\end{figure}

The Quantum Cramer-Rao Bound (QCRB) \cite{Tsang2011,Miao2017} defines the best possible sensitivity of linear sensors across all signal frequencies in the absence of decoherence for a given optical power.
QCRB associates the device's sensitivity to the interaction between the meter and the object, and ultimately to the energy fluctuations of a meter\,\cite{Braginsky2000}.
Attaining QCRB is only possible with a pure quantum state upon interaction\,\cite{braunstein1994statistical, braunstein1996generalized}, with quantum decoherence in any practical setup preventing this, as was investigated in \cite{Dorner2009,Demkowicz-Dobrzanski2013, Demkowicz-Dobrzanski2015,Miao2019}.
However, not all sources of decoherence impact the sensitivity in the same way.
We recognize three distinct loss sources: injection, readout and internal, see Fig.\,\ref{fig_gwsetup}.
Injection loss occurs upon injecting externally squeezed state into the detector cavity and limits the available squeeze factor.
Readout loss impacts the light propagating from the cavity towards the photodetector, including the photodetection process itself.
Internal loss happens during the measurement process itself and directly affects the purity of the state upon interaction with the object.

In this work, we posit that internal loss establishes the fundamental limit for cavity-enhanced interferometric sensing.
In contrast, readout loss can be mitigated by internal squeezing, which requires a second-order non-linear crystal placed inside the interferometer's cavity pumped with second harmonic light\,\cite{Rehbein2005, V.Peano2015, korobko2017beating,korobko2018engineering, korobko2019quantum}.
The pump phase specifies the quadrature phase subject to parametric amplification.
We examine a general case incorporating all three loss sources, and investigate the enhancement in sensitivity as well as sensitivity-bandwidth product\,\cite{korobko2017beating}.
We also demonstrate the benefit of internal squeezing compared to output amplification\,\cite{Caves1981}.

Our limit supports our recent experimental readout loss mitigation via internal squeezing\,\cite{korobko2023main}.
Furthermore, it provides a comprehensive framework whithin which the specific design of cavity-enhanced detectors can be optimised -- for gravitational-wave detection and beyond.

\section{Decoherence-induced limit in a lossy cavity-enhanced detector with internal squeezing}
\begin{figure}
  \vskip0.1cm
\includegraphics[width=1\columnwidth]{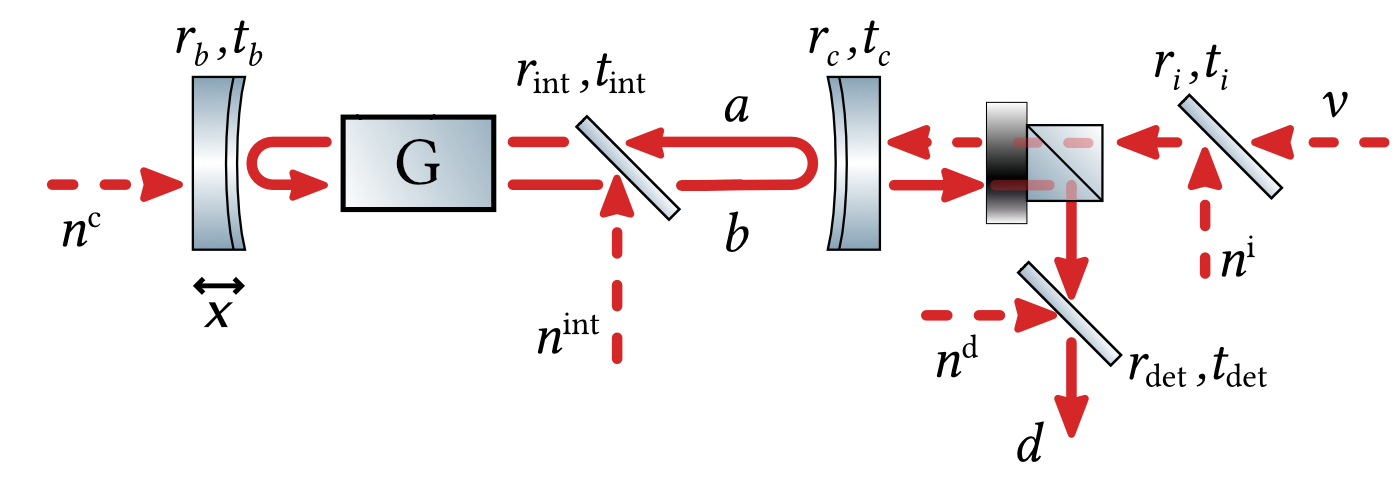}
\caption{Schematic representation of the internal squeezing cavity. 
An external force moves the mirror $x$, producing the phase shift on the light field. 
Signal together with noise are amplified by a factor $G$ inside a nonlinear crystal.
Outgoing field $d$ is separated from the incoming squeezed field $v$.
Optical loss $r_{\rm{i}, \rm{int}, \rm{det}}$ (injection, internal and readout correspondingly) is introduced as a beam-splitter with corresponding amplitude reflectivity.
Optical loss couples in vacuum fields $n^{\rm{c},\rm{int},\rm{i},\rm{det}}$ that destroy coherence of squeezed state.
Fields $a,b$ define counter-propagating quantum fields inside the cavity, field $d$ is detected with a homodyne detector. }\label{fig:sup_simple}
\end{figure}

We model a force sensor as a Fabry-Perot cavity with a mirror that is displaced under an action of an external force, which imprints phase modulation on the reflected light field.
Such a model is equivalent to the full interferometer, as shown in Fig.\,\ref{fig_gwsetup}, in terms of quantum noise\,\cite{Buonanno2003}.
Internal squeezing is produced by a non-linear medium inside the cavity, and squeezed vacuum injected externally, see Fig. \ref{fig:sup_simple}.
In addition, we complement the detector by an output amplifier\,\cite{Caves1981}, which amplifies the signal quadrature by a factor of $\zeta$ (in power) and proportionally deamplifies the orthogonal quadrature.
In our study we focus on the shot-noise limited sensitivity, assuming quantum back-action to be evaded by appropriate measures\,\cite{Kimble2001}.

We follow the input-output formalism\,\cite{Caves1985a, Schumaker1985a,Danilishin2012} for solving the equations of motion for the light fields shown in Fig.\,\ref{fig:sup_simple}. 
We further make a single-mode approximation, assuming the cavity to have high enough finesse to have a well-defined single mode within a free spectral range of the longitudinal resonances.
A signal at the Fourier frequency $\Omega$ couples only to this mode.
The complete derivation can be found in Appendix, where we compute the simplified noise spectral density $S_n(\Omega)$ and the spectral shape of the signal transfer function $\mathcal{T}(\Omega)$:
\begin{align}
  S_{n}(\Omega) =& \frac{1}{\mathcal{D}(\Omega)}\left[\mathcal{D}(\Omega) - 4 q T_c (1-\epsilon_{\rm read}) - \right.\nonumber \\
  &- (1-\beta^{-1})(1-\epsilon_{\rm read})\times\nonumber\\
  &\left.\times \left((q - T_c + \epsilon_{\mathrm{int}})^2 + 16 \Omega^2 \tau^2\right)\right]\label{eq:full_sn}\\
  |\mathcal{T}(\Omega)|^2 =& \frac{8 \pi P_c}{\hbar \lambda c} \frac{4T_c(1-\epsilon_{\rm read})}{\mathcal{D}(\Omega)}\\
  \mathcal{D}(\Omega) =& (q +T_c + \epsilon_{\mathrm{int}})^2 + 16 \Omega^2 \tau^2.\label{eq:full_T},
\end{align}
where $q$ is the gain in a single pass through the crystal, $T_c$ is the power transmission through the front mirror, $\epsilon_{\rm int, read}$ is the internal and readout power loss correspondingly, $\tau$ is a single-pass time inside the cavity, $P_c$ is the total power inside the cavity (which we assume to be fixed), $\lambda$ is the central wavelength, $c$ is the speed of light, and $\beta = e^{2r_{\rm ext}}$, where $r_{\rm ext}$ is the external squeezing rate.
We assumed the output amplification and the injection loss to be zero, and will add them later in the text. 
The derived expression allows to optimize internal squeezing $q$ in order to achieve the highest sensitivity $S_n(\Omega)/|\mathcal{T}(\Omega)|^2$ and study the limiting cases. 

We first consider a low-frequency case: $\Omega \approx 0$, which allows to see the characteristic behavior of different limiting cases better and doesn't affect the conceptual understanding.
We make a reference to a QCRB in a lossless case, $\epsilon_{\mathrm{int}}=0, \epsilon_{\rm read}=0$:
\begin{equation}
  S_{hh}^{\mathrm{QCRB}} = \frac{\hbar \lambda c}{8 \pi P_c L^2} \times\frac{(T_c-q)^2}{\beta T_c},
\end{equation}
where we normalized the sensitivity to the GW strain $h_0 = x/L$.
Notably, in the lossless case the sensitivity becomes unlimited, $S_{hh}^{\mathrm{QCRB}} = 0$, at the parametric threshold: $\gamma_s = \gamma_c$ or at infinite injected squeezing, $1/\beta = 0$.
Optical losses will prevent from achieving this QCRB.
It is therefore possible to compute the new limit $S_{hh}^{\epsilon}$, induced by the losses.
We start by computing the reference sensitivity without internal squeezing, $q=0$, in two limiting cases: when there is no externally injected squeezing and when it is infinitely large:
\begin{align}
  \left.S_{hh}^{\epsilon}\right|_{q=0, \beta=1}& =  \,\frac{\hbar \lambda c}{8 \pi P_c L^2}\frac{1}{1-\epsilon_{\rm read}}\times\nonumber \\
  &\,\times\left(\frac{T_c}{4} \epsilon_{\rm read} + \frac{1}{2}\epsilon_{\mathrm{int}} + \frac{\epsilon_{\mathrm{int}}^2}{4T_c}\right),\\
  \left.S_{hh}^{\epsilon}\right|_{q=0}& \xrightarrow[\beta\rightarrow \infty]{} \,\frac{\hbar \lambda c}{8 \pi P_c L^2}\frac{1}{1-\epsilon_{\rm read}}\times \nonumber\\
    &\,\times\left(\frac{T_c}{4} \epsilon_{\rm read} + \frac{2-\epsilon_{\rm read}}{2}\epsilon_{\mathrm{int}} + \frac{\epsilon_{\rm read}\epsilon_{\mathrm{int}}^2}{4T_c}\right).
\end{align}
Another characteristic reference is the case when internal squeezing is at its threshold value: $q = q^{\rm th} := T_c + \epsilon_{\rm int}$:
\begin{align}
  \left.S_{hh}^{\epsilon}\right|_{\beta=1} = &\frac{\hbar \lambda c}{8 \pi P_c L^2}\frac{1}{1-\epsilon_{\rm read}}\left(T_c \epsilon_{\rm read} + \epsilon_{\mathrm{int}} + \frac{\epsilon_{\mathrm{int}}^2}{4T_c}\right),\\
  S_{hh}^{\epsilon} \xrightarrow[\beta\rightarrow \infty]{} &\frac{\hbar \lambda c}{8 \pi P_c L^2}\frac{1}{1-\epsilon_{\rm read}}\left(T_c \epsilon_{\rm read} + \epsilon_{\mathrm{int}} + \frac{\epsilon_{\mathrm{int}}^2}{4T_c}\right).\label{eq:threshold}
\end{align}
This case has been discussed in Ref.~\cite{Miao2019} as the ultimate detection limit.
It is interesting to note that in the case without internal squeezing the limit is 4 times lower than in the case with at-threshold internal squeezing.
This occurs due to 6\,dB of signal deamplification in the case of internal squeezing at threshold\,\cite{korobko2017beating}.

\begin{figure}[t]
  \centering
  \includegraphics[width=1\linewidth]{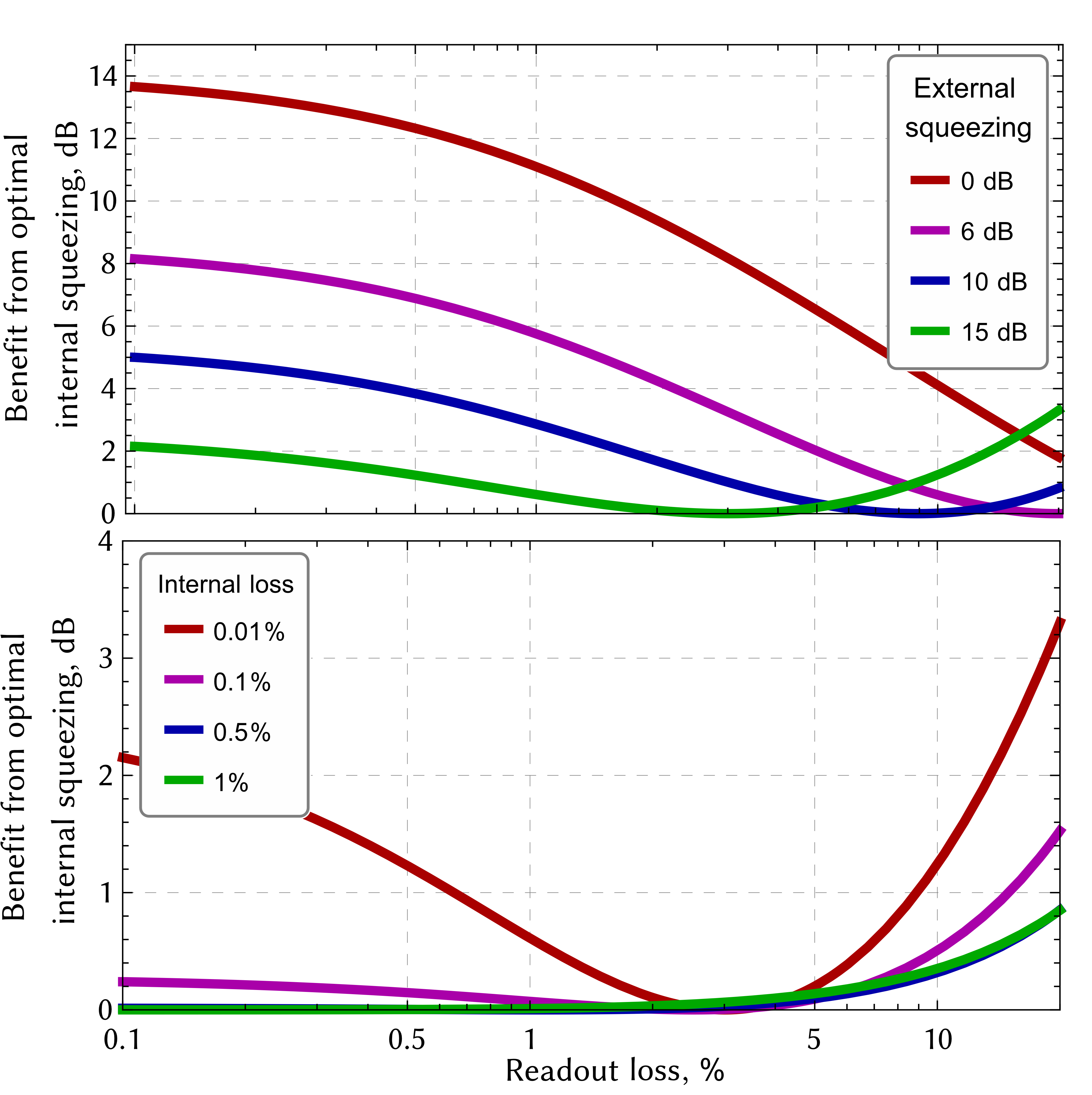}
       \caption[]{Relative sensitivity improvement from optimal internal squeezing compared to the case without internal squeezing.
       Top: relative improvement as a function of readout loss for different levels of external squeezing, for $\epsilon_{\mathrm{int}} = 0.001$. 
       Internal squeezing is almost always beneficial, except for a single value of readout loss for each curve.
       For higher readout loss, it is optimal to amplify inside the interferometer cavity.
       Bottom: relative improvement as a function of readout loss for different levels of internal loss, for 15\,dB external squeezing. 
       For high internal loss, internal squeezing becomes beneficial when the detection loss is high. For low internal loss, internal squeezing is almost always beneficial, except for a single readout loss value.
       For losses higher than that it becomes optimal to amplify inside the cavity.
       For all plots $T_c = 0.01, \Omega=0$.}\label{fig:decoherence_benefit}
\end{figure}

These two references allow comparison with an optimal sensitivity $S_{hh}^{\rm opt}$, for which the optimal internal gain $q_{\mathrm{opt}}$ allows to reach the new sensitivity limit:
\begin{align}
  S_{hh}^{\mathrm{opt}} &= \frac{\hbar \lambda c}{8 \pi P_c L^2}\left(\frac{T_c\epsilon_{\rm read}}{1+\epsilon_{\rm read}(\beta-1)} + \epsilon_{\mathrm{int}}\right),\\
  q^{\rm opt} &= T_c \left(1 - \frac{2 \epsilon_{\rm read}}{\beta (1-\epsilon_{\rm read}) - \epsilon_{\rm read}}\right) - \epsilon_{\rm int}.\label{eq:optimal_gain}
\end{align}
This limit constitutes the ultimate limit on the sensitivity, and overcomes the limit in Ref.\,\cite{Miao2019}.
Note, that in this case the QCRB is not zero, since the internal squeezer operates below threshold.
In \figref{decoherence_benefit} we show the benefit from optimal internal squeezing compared to the case without internal squeezing for different values of losses and externally injected squeezing.
One can see that the benefit is higher for lower external squeezing, since then internal squeezing works to generate more squeezing.
For higher external squeezing, the benefit increases with increased readout loss as it is compensated by internal squeezing.
Internal loss defines the sensitivity limit, so the higher it is, the smaller is the benefit from internal squeezing.
There also exists a readout loss value for which it's optimal not to squeeze internally, since the introduced signal deamplification exactly cancels a potential benefit.

In the absence of external squeezing we achieve:
\begin{equation}
  \left.S_{hh}^{\mathrm{opt}}\right|_{\beta=1} = \frac{\hbar \lambda c}{8 \pi P_c L^2}\left(T_c\epsilon_{\rm read} + \epsilon_{\mathrm{int}}\right)
\end{equation}
which overcomes the limit for the at-threshold operation.
The optimal gain here is below the threshold
\begin{equation}
  \left.q_{\mathrm{opt}}\right|_{\beta=1} = q_{\mathrm{th}} - 2(T_c\epsilon_{\rm read}+\epsilon_{\mathrm{int}}).
\end{equation}
In this case the purpose of internal squeezing is to generate as much squeezing as possible, at the same time keeping the deamplification of the signal as low as possible, taking into consideration the losses in the system.
This regime was explored by us experimentally before in Ref.\,\cite{korobko2017beating}.

\begin{figure}[t]
  \centering
   \begin{minipage}{1\columnwidth}
     \centering
          \includegraphics[width=1\columnwidth]{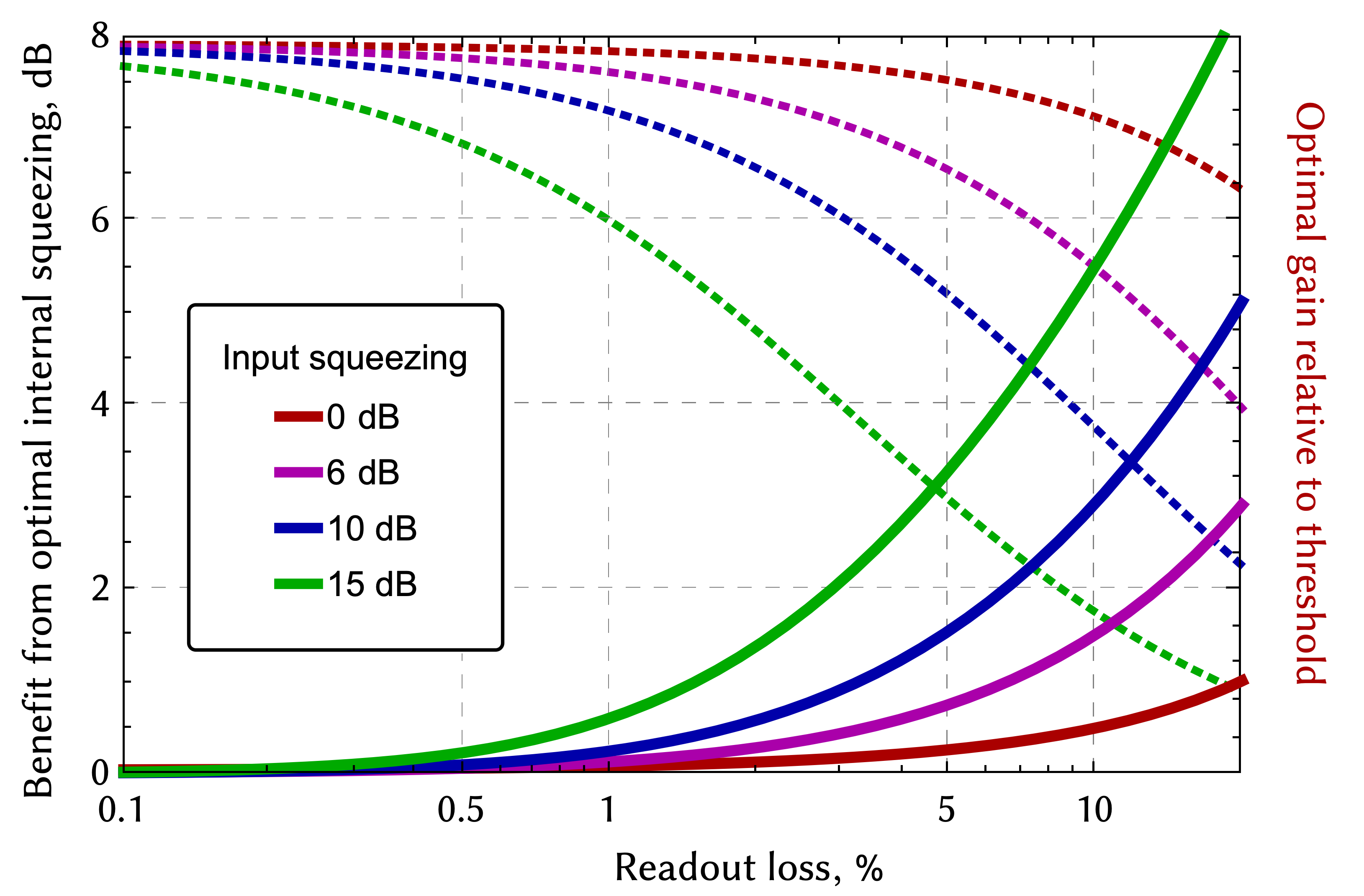}
   \end{minipage}
   \begin{minipage}{1\columnwidth}
     \centering
          \includegraphics[width=1\columnwidth]{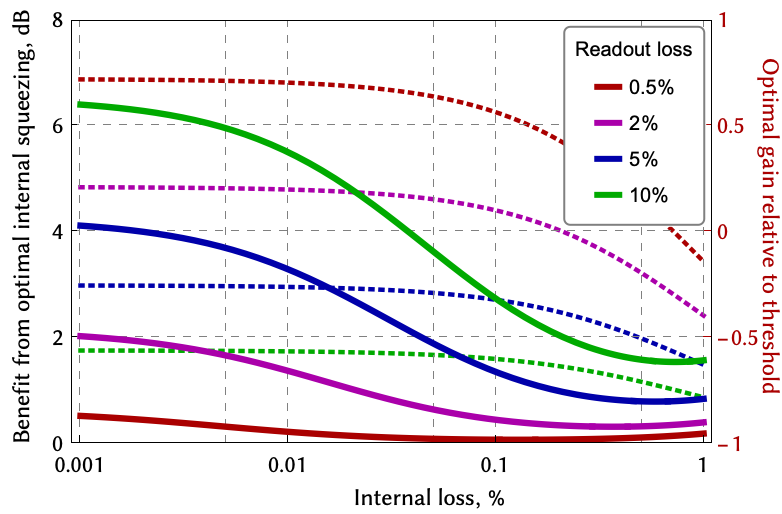}
   \end{minipage}
            \caption[]{Relative sensitivity improvement from optimal internal squeezing compared to the case of internal squeezing at threshold: the improvement $S_{hh}^{\mathrm{opt}}/S_{hh}^{\epsilon}$ (solid), and the corresponding optimal internal gain relative to the threshold $q_{\mathrm{opt}}/q_{\mathrm{th}}$ (dashed).
             (Top) relative improvement as a function of readout loss for different levels of external squeezing, for $\epsilon_{\mathrm{int}} = 0.001$;
             (bottom) relative improvement as a function of internal loss for different levels of readout loss, for 15 dB external squeezing.
             Internal gain is optimized to squeezing for low losses, but for high losses switches to parametric amplification regime (values below zero).
             The plots demonstrate that optimised internal squeezing is always beneficial and leads to an improvement compared to the limit presented in Ref.\,\cite{Miao2019}.
              For all plots $T_c = 0.01, \Omega=0$.}\label{fig:decoherence_benefit_gain}
\end{figure}

Finally, we can compute the ultimate limit for infinite squeezing, which is achieved by maximally amplifying the signal quadrature inside the cavity  $q = -q^{\rm th}$:
\begin{equation}\label{eq:decoherence_limit}
  S_{hh}^{\mathrm{opt}} \xrightarrow[\beta\rightarrow \infty]{}  \frac{\hbar \lambda c}{8 \pi P_c L^2}\epsilon_{\mathrm{int}}.
\end{equation}
This is the ultimate limit to the sensitivity of a linear detector, and it can be significantly lower than the sensitivity at threshold\,\eqref{eq:threshold}.
In \figref{decoherence_benefit_gain} we show the benefit from optimal internal squeezing compared to the on-threshold case.
As expected from the equations, when the readout loss is small, the benefit is negligible (since internal squeezing works best when it can compensate the readout loss).
However, as the readout loss and externally injected squeezing increase, the benefit increases too, and at the same time it requires generating less squeezing inside, quickly transitioning into the amplification regime (negative values for $q$).
For example, already for a readout loss of around 5\% (which is smaller than the current loss in gravitational-wave detectors) it becomes optimal to amplify inside (for 15\,dB of external squeezing).

\subsection{Sensitivity-bandwidth relation}
In the previous section we investigated only the effect of internal squeezing on the peak sensitivity.
It is important to also consider the detection bandwidth and the sensitivity-bandwidth product as a generic measure\,\cite{korobko2017beating}.
Internal squeezing reduces the effective bandwidth of the system, and the higher the squeezing is, the smaller is the bandwidth.
However, since the peak sensitivity increases faster, the overall sensitivity-bandwidth product is enhanced beyond the standard sensitivity-bandwidth limit (SSBL) for coherent light.
Eqs.\,\eqref{eq:full_sn},\eqref{eq:full_T} allow to analyse the sensitivity-bandwidth product by defining the bandwidth $\mathcal{B}$ as the half-width half-maximum of the sensitivity. 
While the full expressions for the bandwidth are rather bulky, we can analyse them numerically.

\begin{figure}[hbt!]
\includegraphics[width=1\columnwidth]{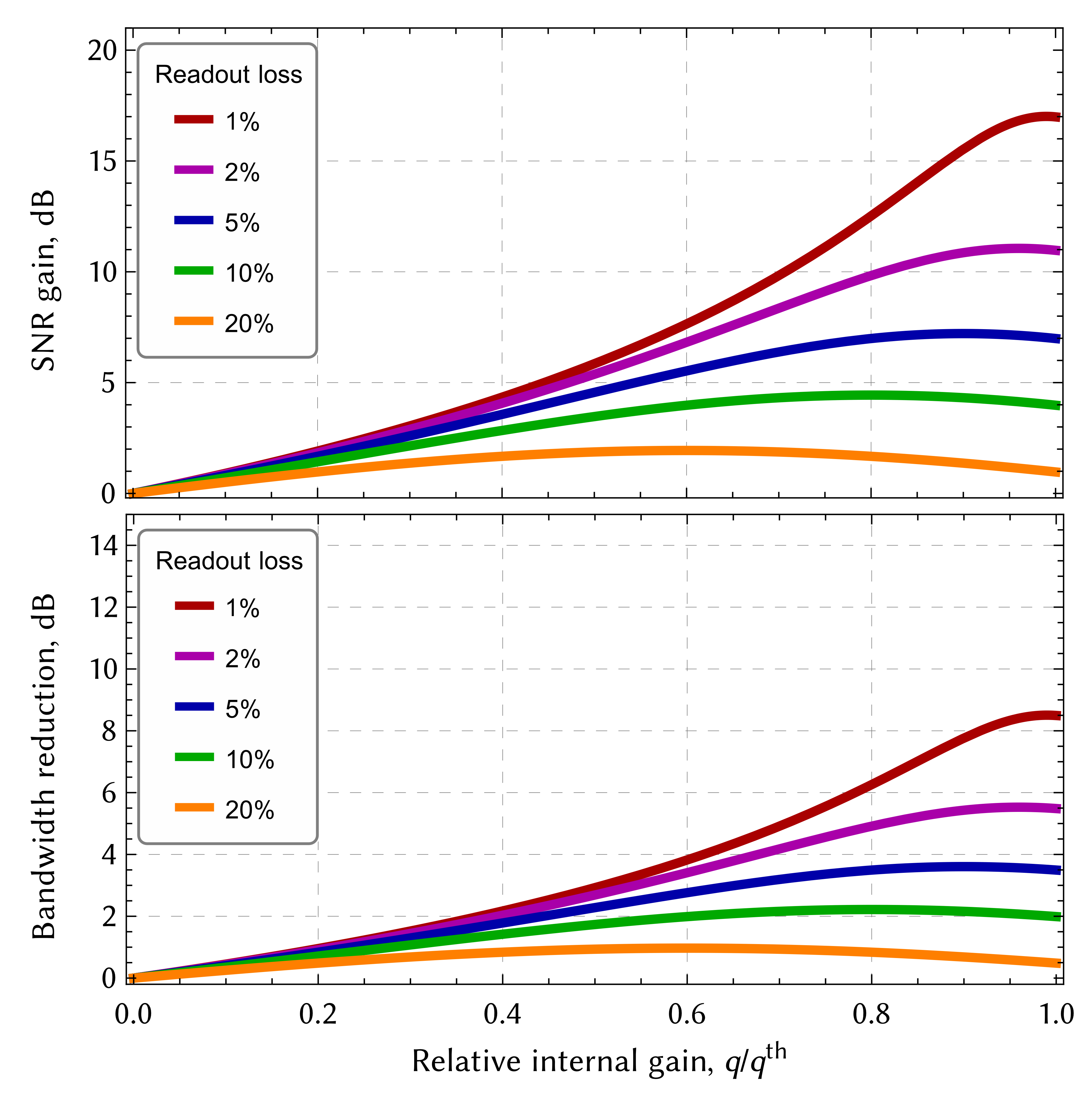}
\vskip-10pt\caption[]{Internal squeezing operating without external squeezing. 
The noise is reduced due to the produced squeezing, and the signal is deamplified at the same time.
The SNR gain (top) and the reduction in the bandwidth (bottom) depends on the internal gain. 
For different values of readout loss, there exists an optimal internal gain for which the SNR is maximized.
Although the bandwidth is reduced, the sensitivity-bandwidth product is still increased, as we demonstrated experimentally in Ref.\,\cite{korobko2017beating}. 
In both cases $T_c=0.01, \epsilon_{\rm int}=0$.
}\label{fig:decoherence:sup_theory}
\end{figure}

In Fig.~\ref{fig:decoherence:sup_theory}, we show the gain in the SNR relative to the case without internal squeezing, and the corresponding reduction in the squeezer's bandwidth, when no externally injected squeezing is present.
As in the previous Section, here the clear optimum point is reached for every internal gain value.

\begin{figure}[hbt!]
\includegraphics[width=\columnwidth]{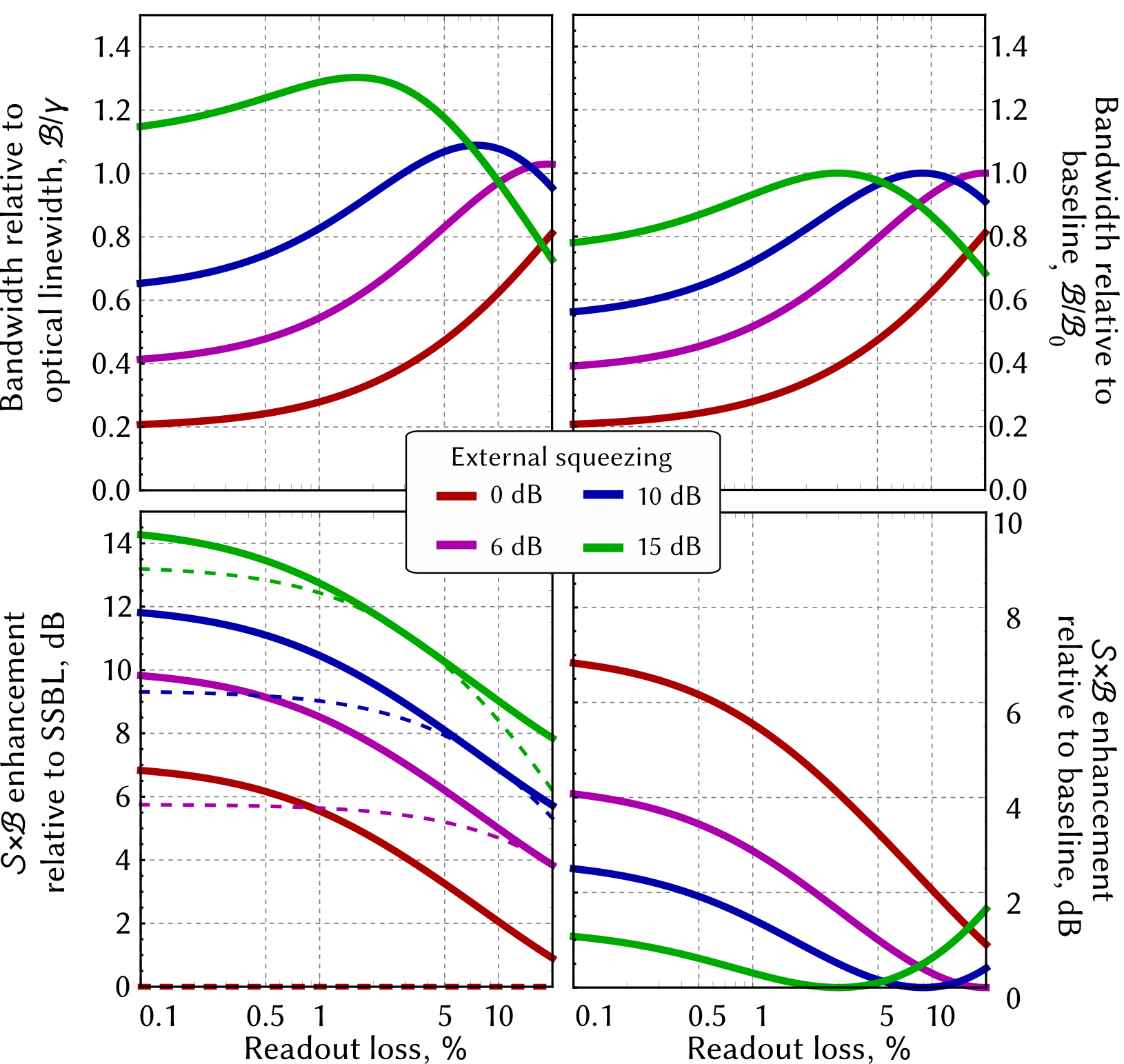}
\caption[Effect of optimal internal squeezing on the detection bandwidth and the sensitivity-bandwidth product.]{Effect of optimal internal squeezing on the detection bandwidth (top) and the peak sensitivity-bandwidth product $\mathcal{S}\times\mathcal{B}$ (bottom).
Left: the changes relative to the standard interferometer without internal or external squeezing.
Right: the changes relative to the baseline interferometer without internal squeezing.
The optimal internal squeezing is different for different parameters, and influences the bandwidth differently.
The sensitivity-bandwidth product is enhanced always except for a single specific value of readout loss.
Bottom left: the solid line is a combination of optimal internal squeezing with external squeezing, dashed line is only external squeezing.
external squeezing provides the main contribution to the enhancement, internal squeezing helps to suppress the readout losses.
For all plots $T_c = 0.01, \epsilon_{\mathrm{int}} = 0.0001$.}\label{fig:decoherence:ssb}
\end{figure}

When external squeezing is injected, internal squeezing acts in different ways depending on the loss level, as described in the previous Section.
The bandwidth is always smaller than in the case without internal squeezing, but its dependence on the losses is nontrivial.
In Fig.\,\ref{fig:decoherence:ssb}, we show the enhancement relative to the case with external squeezing, demonstrating that the sensitivity can be always enhanced, despite the reduction in bandwidth.
We also show the enhancement relative to the SSBL, demonstrating that although the main contribution comes from the external squeezing, internal squeezing helps to gain sensitivity.
In this case the bandwidth can also be enhanced in some parameter regimes, due to nontrivial dependence of the bandwidth on losses and squeezing levels.

\section{Injection loss and output amplification}
Finally, in order to complete the consideration, we provide the expression for the sensitivity including the injection loss for external squeezing as well as an additional amplifier before the detection, as it was first proposed by Caves\,\cite{Caves1981}.
Such three-stage squeezing scheme (input, internal, output) would yield the best sensitivity with any combination of losses.

In order to include the injection loss $\epsilon_{i}$, we add a loss term when computing the spectral density, \emph{i.e.} $S_{vv} = \beta (1-\epsilon_{i}) + \epsilon_{i}$.
The output amplification would be represented as a linear gain $\zeta$ just before the detection loss term.

The resulting sensitivity with optimized internal gain (in the single-mode approximation):
\begin{multline}\label{eq:decoherence:full}
  S_{hh}^{\mathrm{opt}}(\Omega) = \frac{\hbar \lambda c}{8 \pi P_c L^2}\times\\
  \times\left[\frac{T_c \epsilon_{\mathrm{det}}\left(1-\epsilon_{i}(1 + \beta)\right)}{\beta \epsilon_{\rm read} + \zeta(1-\epsilon_{\rm read})(1-(1-\beta)\epsilon_{i})}+ \right. \\ + \epsilon_{\mathrm{int}} 
 + \frac{4\Omega^2 \tau^2}{T_c \beta \zeta \left(1-\epsilon_{\mathrm{det}}\right)} \left(\beta\epsilon_{\mathrm{det}} + \right. \\ \biggl. \left. + \zeta(1-\epsilon_{\rm read})(1-(1-\beta)\epsilon_{i})\right)\biggr].
\end{multline}
It is interesting to consider limiting cases here as well.
In the limit of infinite external squeezing (at zero frequency):
\begin{equation}\label{eq:limit_injection}
  S_{hh}^{\mathrm{opt}}(0) \xrightarrow[\beta\rightarrow \infty]{} \frac{\hbar \lambda c}{8 \pi P_c L^2}\left[\frac{T_c \epsilon_{\mathrm{det}} \epsilon_{i}}{\epsilon_{\mathrm{det}} + \zeta\epsilon_{i}\left(1-\epsilon_{\mathrm{det}}\right)} + \epsilon_{\mathrm{int}}\right].
\end{equation}
Without output amplification ($\zeta=1$), the limit is dependent on the injection loss.
The injected state is no longer pure, and that impurity contributes to the interaction Hamiltonian, thus preventing from achieving the QCRB, as discussed in the introduction.
Internal squeezing is unable to compensate that effect.
Injection loss effectively reduces the level of external squeezing, and the state can be kept rather pure by optimizing the external squeezing strength.
An additional effect of destroyed correlations between two quadratures of the field used for back-action evasion, will ultimately limit the sensitivity in back-action-dominated region, which we do not consider here.

Interestingly, in the limiting case of infinite output amplification, the result is independent on the external squeezing $\zeta\rightarrow\infty$:
\begin{equation}
  S_{hh}^{\mathrm{opt}}(0) \xrightarrow[\zeta\rightarrow \infty]{} \frac{\hbar \lambda c}{8 \pi P_c L^2}\epsilon_{\rm int}, \quad q_{\mathrm{opt}} = (T_c - \epsilon_{\mathrm{int}}).
\end{equation}
Here, the readout loss is fully compensated by output amplification, and internal squeezing can be used to create maximal squeezing (taking into account balance between signal deamplification and internal loss).

Finally, we can compare the limits from the internal squeezing and output amplification.
In essence, they achieve the same thing: they allow to evade the readout loss.
Output amplification without internal squeezing reaches the limit:
\begin{equation}
  \left.S_{hh}(0)\right|_{q=0} \xrightarrow[\zeta\rightarrow \infty, \beta\rightarrow \infty]{}  \frac{\hbar \lambda c}{8 \pi P_c L^2}\left(\epsilon_{\rm int} + \frac{\epsilon_i(T_c - \epsilon_{\rm int})^2}{4T_c}\right).
\end{equation}
Comparing this expression with Eq.\eqref{eq:limit_injection} for $\zeta=1$ we will find that they both reach the same limit if there's no injection loss.
However, if injection loss is present, these expressions are different, and in different regimes either one or the other work better.
That does not mean, however, that output amplification is sufficient to reach the ultimate limit. 
Here we considered a perfect output amplification with no added losses, while internal squeezing is considered imperfect (i.e. taking into account the losses).
This added loss would lead to an increased decoherence-induced limit, higher than that of internal squeezing.
The specific implementation of the output amplifier requires a separate study.

\section{Discussion and conclusion}

Internal squeezing mitigates the decoherence effect of optical loss, establishing a generally improved sensitivity limit. Uniquely, internal squeezing enhances sensitivity in two ways: either by directly maximizing the SNR when the loss is minimal, or by amplifying both the signal and the noise to evade significant readout losses. 
In this work, we offer a detailed exploration of various regimes for the application of internal squeezing within realistic parameters.

We also compared internal squeezing with the ideal output amplification\,\cite{Caves1981}, which was recently explored in other contexts\,\,\cite{colomboTimereversalbasedQuantumMetrology2022, manceauDetectionLossTolerant2017, ouEnhancementPhasemeasurementSensitivity2012, knyazevOvercomingInefficientDetection2019,frascellaOvercomingDetectionLoss2021}.
By highlighting the similarities and differences between internal squeezing and this approach, we delineate the limits each can achieve independently and present an optimal design combining both to counteract the three primary loss sources.
Our argument suggests that in the presence of input, internal, and readout loss, the optimal configuration incorporates both output amplification and internal squeezing.
Internal squeezing usually would suffice for low injection loss, but the optimal configuration would be determined by the full parameter set.

In our study, we assumed that such an output amplifier would introduce no loss -- an unrealistic assumption.
Practically, with the presence of an output amplifier, the readout loss must be divided into three stages: before, inside, and after the amplifier.
In the GW detectors -- the primary target for such designs --  the major source of readout loss would occur before the output amplifier: on the Faraday isolator used to inject squeezed light, on the mode-matching to the output mode-cleaner used to purify the optical mode leaving the interferometer, and inside this mode-cleaner.
This scenario limits the utility of output amplification as it would only effectively compensate for photodetection inefficiency, which is typically very low.
Contrarily, internal squeezing can compensate for all these decoherence sources.

In addition to the optical loss, other decoherence mechanisms are present in realistic detectors: phase noise\,\cite{Schnabel2017,kijbunchoo2020low}, dephasing\,\cite{kwee2014decoherence}, mode mismatch\,\cite{McCuller2021response}, and others. 
The impact of these sources on the performance of internal squeezing and the fundamental limits will be the direction of future research.

Internal squeezing is an exceptionally versatile tool for enhancing the sensitivity of various sensors.
It offers manipulation of the sensitivity\,\cite{korobko2017beating}, the bandwidth\,\cite{korobko2019quantum}, or even modifies the dynamics of test masses in optical cavities\,\cite{korobko2018engineering}. 
This paper introduces a new facet to the study of internal squeezing, demonstrating its capability to achieve the fundamental limit imposed by optical decoherence.
To complement this theoretical study, we experimentally demonstrated mitigation of decoherence and optimal internal squeezing in Ref.\,\cite{korobko2023main}.
Together with the current paper, it paves the way for practical implementation of internal squeezing in quantum sensing and underlines its advantages in various practical scenarios.

\textit{Acknowledgements} We thank F.Khalili and H.Miao for insightful discussions. This work was supported by the Deutsche Forschungsgemeinschaft (DFG) under Germany's Excellence Strategy EXC 2121 ``Quantum Universe''-390833306. 

\appendix*
\section{Appendix A: Derivation of spectral density}\label{app:app_a}

Using the perturbation theory, we decompose the light field into a steady-state amplitude with amplitude $A_0$ and laser carrier frequency $\omega_0$ and a slowly varying noise amplitude $a(t)$ (see details in \cite{Danilishin2012}):
  
  \begin{align}
  &A(t) = \sqrt{\frac{2\pi \hbar \omega_0}{\mathcal{A} c}} \left[ A_0 e^{-i\omega_0 t} + a(t) e^{-i\omega_0 t} \right] + {\rm h.c.}\\
  &a(t) = \int_{-\infty}^{\infty} a(\omega_0 + \Omega)e^{-i\Omega t} \frac{d \Omega}{2 \pi},
  \end{align}
  where $\mathcal{A}$ is the laser beam cross-section area and $\hbar$ is the reduced Plank constant. Note that we omit the hats on the operator for brevity, although all the fields are quantised.  
  Since we are interested in the steady-state solutions for the fields, we consider only the noise fields in the frequency domain. 
  We further define the two-photon amplitude and phase quadratures at a sideband frequency $\Omega$ correspondingly:
  \begin{align}
  & a_x(\Omega) = \frac{a(\omega_0 + \Omega) + a^\dag(\omega_0 - \Omega)}{\sqrt{2}},\\
  & a_y(\Omega) = \frac{a(\omega_0 + \Omega) - a^\dag(\omega_0 - \Omega)}{i\sqrt{2}}.
  \end{align}
  These operators obey the commutation relation:
  \begin{align}
  &[a_x(\Omega), a_x(\Omega')] = [a_y(\Omega), a_y(\Omega')] = 0,\\
   &[a_x(\Omega), a_y(\Omega')] = [a_x(\Omega), a_y(\Omega')] = 2 \pi i \delta(\Omega + \Omega').
  \end{align}
  Using these two-photon quadratures, we can apply the input-output formalism \cite{Caves1985a, Schumaker1985a} and find the steady-state fields in the system.

 The signal $s$ appears only in the equations for the phase quadrature of the light field.
 We model the optical loss by a beamsplitter reflecting some part of the light fields to the environment and mixing in some vacuum from the environment. 
The optical parametrical amplification process is not included in the model, we treat crystal as a gain media that linearly amplifies with a gain $G$ a certain quadrature (amplitude in our case) and deamplifies the orthogonal to it; we call $q$ a squeezing factor in a single pass through the crystal.
We also consider the output amplification of the field before detection\,\cite{Caves1981}, which amplifies the signal quadrature by a factor of $\zeta$ (in power) and proportionally deamplifies the orthogonal quadrature.

We start by deriving the full input-output relations for the fields in the process, as shown in \figref{sup_simple}.
\begin{align}
&\begin{cases}
  b^x = &a^x e^{2i\Omega\tau} G^2 r_b t_{\rm int} + n_c^x t_b t_{\rm int} G e^{i\Omega\tau} + \\ & + r_{\rm int} n_{\rm int}^x,\\
  a^x = &r_c b^x + t_c t_i v^x + t_c r_i n_i^xy, \\
  d^x = &\zeta^{-1/2}t_b(t_c b^x - r_c t_i v^x - r_c r_i n_i^x) + r_d n_d^x,
\end{cases}\\
&\begin{cases}
  b^y = &a^y e^{2i\Omega\tau} G^{-2} r_b t_{\rm int} + n_c^y t_b t_{\rm int} G^{-1} e^{i\Omega\tau}  + \\ & + r_{\rm int} n_{\rm int}^y + s t_{\rm int} G^{-1}e^{i\Omega\tau},\\
  a^y = &r_c b^y + t_c t_i v^y + t_c r_i n_i^y, \\
  d^y = &\zeta^{1/2}t_b(t_c b^y - r_c t_i v^y - r_c r_i n_i^y) + r_d n_d^y,
  \end{cases}
\end{align}
where $r_{c, b, \rm{int}, i}$ are the amplitude reflectivity of the front mirror, back mirror, internal loss and injection loss; $t=\sqrt{1-r^2}$ are the corresponding transmissivities; $\tau$ is the cavity single-pass time.

Then, solving equations for the output field, we arrive at:
\begin{align}
  d^x =& \frac{\zeta^{-1/2}t_d}{G^{-2} - r_c r_b t_{\rm int} e^{2i\Omega\tau}} \left[t_c r_{\rm int} n_{\rm int}^x G^{-2} + \right.\\
  &+ n_i^x\left(-r_c r_i G^{-2} + r_b r_i t_{\rm int} e^{2i\Omega\tau}\right) + \\
&\left. + v^x \left(-r_c t_i G^{-2} + r_b t_i t_{\rm int} e^{2i\Omega\tau} \right)\right]  + n_e^x r_d, \\
  d^y =& \frac{\zeta^{1/2} t_d}{G^2 - r_c r_b t_{\rm int} e^{2i\Omega\tau}} \left[s t_c t_{\rm int} e^{q+i\Omega\tau} \right. \\ 
  & + t_c r_{\rm int} n_{\rm int}^y G^2 + \\
  &\left.+ n_i^y\left(-r_c r_i G^2 + r_b r_i t_{\rm int} e^{2i\Omega\tau}\right) + \right. \\
&\left. + v^y \left(-r_c t_i G^2 + r_b t_i t_{\rm int} e^{2i\Omega\tau} \right)\right]  +  n_e^y r_d.
\end{align}
Solving the equations for amplitude quadrature and maximizing the field strength inside the cavity, we arrive at the threshold value:
\begin{equation}
  e^{-q^{\rm th}} = (r_b r_c \sqrt{1 - \epsilon_{\rm int}}).
\end{equation}
We obtain the spectral shape of transfer function $\mathcal{T}(\Omega)$ for the signal $s$, where we introduce power transmissivities: $T_n = t_n^2, R_n = r_n^2$ and losses $\epsilon_n = r_n^2$:
\begin{align}
  |\mathcal{T}(\Omega)|^2 = & \frac{\zeta G^2 T_c (1-\epsilon_d)(1-\epsilon_{\rm int})}{|\mathcal{D}(\Omega)|^2},\\
  |\mathcal{D}(\Omega)|^2 = & G^4 + R_b R_c (1-\epsilon_{\rm int}) - \nonumber\\
   - & 2G^2\sqrt{R_b R_c (1-\epsilon_{\rm int})}\cos2\Omega\tau.
\end{align} 
For computing the noise spectral density, we note that external squeezing is injected into the cavity, with the spectral density $S_{vv} = \beta = e^{2r_{\rm ext}}$, where $r_{\rm ext}$ is the external squeezing rate.
The phase of external squeezing is set to squeezing or anti-squeezing, since any other angle would introduce non-trivial dynamics to the system, which we explored before in a loss-less case in Ref.\,\cite{korobko2018engineering}, while a full treatment requires a separate investigation.
This results the noise spectral density:
\begin{align}
  S_{n}(\Omega) =& 1 - \frac{(1-\epsilon_{\rm read})}{|\mathcal{D}(\Omega)|^2}\left(S_1 \epsilon_{\rm int} + S_2 + \right. \nonumber \\
  & \left.+ S_3 \sqrt{R_b R_c \left(1-\epsilon_{\rm int}\right)} \cos 2\Omega\tau\right)\\
  S_1(\Omega) =& -R_b R_c + \zeta\left(G^2 T_c (T_b - G^2) + \right. \nonumber\\
  &  \left.+ R_b \beta^{-1} (1-\epsilon_i) + R_b \epsilon_i\right)\\
  S_2(\Omega) =& G^4 + R_b R_c - \zeta\left[\left(\epsilon_i + \beta^{-1}\left(1-\epsilon_i\right)\right)\times\right.\nonumber\\
  & \left. \times \left(R_b - G^4R_c\right) + G^2T_b T_c\right]\\
  S_3(\Omega) =& 2 G^2\left(-1 + \zeta\beta^{-1}\left(1-(1-\beta)\epsilon_i\right)\right)
\end{align}
The sensitivity is defined as a noise-to-signal ratio: $S_n(\Omega)/|\mathcal{T}(\Omega)|^2$.

The expression for the spectral density and signal transfer function can be simplified by making the standard single-mode approximation.
The amplitude transmissivities of the coupling and the back mirrors, as well as the internal loss, are much smaller than unity, and we can approximate correspondingly $R_{\mathrm{c}}\approx 1 - T_c/2$, and set $T_b=0$, effectively including into the internal loss.
The single-pass gain $G$ is close to unity, so we can approximate $G\approx1+q/4$, where $q$ is the double-pass power gain factor; the frequency of interest is much smaller than the FSR of the cavity $\Omega \ll 1/2\tau$, which enables us to make a Taylor expansion: $\cos \Omega \tau \approx 1- \Omega^2 \tau^2/2$.
The signal $s$ is due to the modulation adds a phase shift on the light reflected of the movable mirror: $e^{2 i k x(\Omega)}\approx 1 + 2 i k_p x(\Omega)$, where $k_p$ is the light wave vector, and $x(\Omega)$ is a small mirror displacement. The signal is then represented as $s(\Omega) = \sqrt{8 \pi P_c (\hbar \lambda c)^{-1}} x(\Omega)$.
We further posit, that the average light power inside remains fixed and does not get amplified by the parametric process.
Finally, we first study the case without injection loss, $\epsilon_{i}=0$, and 
the output amplification, $\zeta=1$.

With these assumptions we achieve the full expression for the noise spectral density and the signal transfer function in the single-mode approximation:
\begin{align}
  S_{n}(\Omega) =& \frac{1}{\mathcal{D}(\Omega)}\left[\mathcal{D}(\Omega) - 4 q T_c (1-\epsilon_{\rm read}) - \right.\nonumber \\
  &- (1-\beta^{-1})(1-\epsilon_{\rm read})\times\nonumber\\
  &\left.\times \left((q - T_c + \epsilon_{\mathrm{int}})^2 + 16 \Omega^2 \tau^2\right)\right]\label{eq:full_sn}\\
  |\mathcal{T}(\Omega)|^2 =& \frac{8 \pi P_c}{\hbar \lambda c} \frac{4T_c(1-\epsilon_{\rm read})}{\mathcal{D}(\Omega)}\\
  \mathcal{D}(\Omega) =& (q +T_c + \epsilon_{\mathrm{int}})^2 + 16 \Omega^2 \tau^2.\label{eq:full_T}
\end{align}

\bibliography{bibliographyCompensating}
\end{document}